\documentstyle[aps,amstex,amssymb,times,fancybox]{revtex}
\tighten
\newtheorem{Cor}{Corollary}
\newtheorem{Proposition}[Cor]{Proposition}
\begin{document}
\draft
\title{\textit{Principia Physica}}
\author{Mukul Patel}
\address{$<$math@@mailroom.com$>$\\ 
principia.webjump.com\\
16389 Saratoga Street, San Leandro, CA 94578,  USA\\
\textbf{(Dedicated to Zoya)}}
\date{August 5, 2000}
\maketitle
\boldmath
\begin{abstract}
A comprehensive physical theory \emph{explains} all aspects of the physical 
universe, including quantum aspects, classical aspects, relativistic aspects,
their relationships, and unification. The central \textit{nonlocality principle}
leads
to a \textit{nonlocal geometry} that explains entire quantum phenomenology,
including two-slit experiment, Aspect-type experiments, quantum randomness,
tunneling etc. The infinitesimal aspect of this geometry is a usual
(differential) geometry, various aspects of which are energy-momentum, 
spin-helicity, electric, color and flavor charges. Their interactions are 
governed by a mathematically automatic field equation---also a grand 
conservation principle. New predictions: a new particle property;
bending-of-light estimates refined over relativity's; shape of the universe; a
\emph{no gravitational singularity} theorem; etc. Nonlocal physics is 
formulated using a
\textit{nonlocal calculus} and \textit{nonlocal differential equations,} 
replacing inadequate 
local concepts of Newton's calculus and partial differential equations. Usual
quantum formalisms follow from our theory---the latter doesn't rest on the
former.
\end{abstract}
\pacs{(PACS: 03, 11, 04, 12)}

\setcounter{secnumdepth}{4}
\setcounter{tocdepth}{4}

\tableofcontents
\setcounter{section}{-1}
\section{\textbf{Introduction}}
The primary purpose of a scientific theory is to understand complex phenomena
with the aid of simpler and readily comprehensible concepts. Modern physics is
faced with a great challenge---quantum phenomena.
These still lack consistent rational explanation after a hundred years since
their discovery. While there are many formalisms in use to compute numbers, there is hardly any
comprehensible explanation of the phenomena.
Most of the paradoxes of the 
quantum theory are paradoxes of 
the theory rather than those of observed phenomena. Clearly, a fresh 
consistent set of
concepts are needed to actually understand the seemingly 
bizarre 
quantum phenomena.
Thus, we abandon the entire opaque machinery of quantum mathematics and all its
interpretations. 
Aspect-type experiments reveal the inherent nonlocality
of the physical world.
Hence, we also abandon the very basis of classical physics---the tacit 
assumption
that phenomena are governed by local mechanisms. Instead, we propose a 
nonlocal physics, constructed
from scratch. This
physics also lays bare the integrated reality which underlies all 
the myriad 
fields, particles, their properties, and their fields. 

Fortunately, 
this entire new conceptual framework can 
be 
deduced from one physical principle, the \textit{nonlocality principle}. 
The latter leads to a 
nonlocal geometry, which is specified by a nonlocal connection (as opposed to
classical, local connection in the sense of differential geometry). This 
nonlocal connection explains entire quantum phenomenology on one hand while 
its local aspect, a classical connection, is the universal field which
yields  an integrated geometric description of all the forces, particles, 
fields, energy-momentum, charges, and other quantum numbers.

Due to nonlocality, fields can't be assumed to be smooth or even continuous. 
Thus, the local concepts of Newton's calculus are inadequate to 
\emph{completely}
describe the physical world, and predict outcomes of experiments. While the 
classical physics is encoded in terms of relationships among (local) rates 
of change of physical
quantities, the crucial concepts of nonlocal physics 
involve the 
way physical
quantities are related to each other nonlocally. Thus,
the new laws should be formulated as statements of these nonlocal 
relationships. For these reasons, we devise a new 
\emph{nonlocal} calculus and nonlocal differential equations.

Although this calculus correctly, and completely, encodes nonlocal dependences
among
fields, the fields are still local in that they are
defined point-wise. The nonlocal
connection mentioned above is not a field defined point-wise. It is
defined at \emph{ordered pairs} of points; and its value
at a pair relates vectors and tensors at one
point with those at the other. To analyze this essentially nonlocal
field and its various aspects, we devise a
calculus of nonlocal fields, along with a nonlocal differential geometry. 
Then we
write down \emph{the} field equation, in terms of concepts of this
calculus. This equation governs all aspects of the
physical world.

We make no 
attempt to explain any of the current quantum formalisms, nor is any
consideration given to the paradoxes arising out of these formalisms. We
only explain observed physical phenomena. Our theory is formulated completely 
using real numbers---noncommutative variables are not needed. 

Besides explaining a great many unexplained phenomena, several new predictions
are deduced.

It can't be over-emphasized that, while current theoretical trends in science and philosophy actively 
shun determinism and rationality, 
our theory brings us back to
the realm of classical logic, and to a determinism stronger than that of
Newton's. Ironically enough, this form of determinism
has plenty of room for `free will', and it also causes the apparent quantum
randomness.

This report consists of three chapters. Chapter \ref{theprinciple} 
introduces 
the nonlocality 
principle and explains quantum phenomena. Chapter \ref{field} examines the fundamental field, which follows from
the nonlocality principle. Since this field is a nonlocal object and has a local,
infinitesimal aspect, the analysis takes three forms. The first analyzes
the local aspect using infinitesimal methods; i.e., using Newton's calculus.
Here, various aspects of the local aspect will be identified with
various properties of particles. Using these fields we can build various 
particles, and the field equation mandates that they
should interact.
This local analysis works only for fields measured at larger scales, and
when measured at smaller scales, the effects of nonlocality manifest, we lose local
smooth nature of fields. Consequently,
we can't even formulate laws governing these using local calculus, and we can't predict events. Consequently, we next
analyze the local aspect using a nonlocal analysis. This yields a more natural
set of equations encoding complete information on the field.
We are still left with an unknown---the nonlocal connection
itself. This being a truly nonlocal object, we devise a calculus for such
objects. For more details, see the table of contents. Chapter \ref{conclusion}
lists several
new predictions. We have systematically suppressed much technical details.
An exhaustive discussion of concepts and technical details, can be found in the
forthcoming research monograph, \textit{Principia Physica,} by the author.

\setcounter{section}{0}
\section{THE PRINCIPLE}\label{theprinciple}
\subsection{\textbf{Nonlocal field}}
\subsubsection{Nonlocality Principle}\label{principle}
Classically, it is conceived that an individual event affects events only in
its immediate vicinity, and this effect travels from point to
point with definite speed. The discovery of a series of quantum
effects, which culminated in Aspect-type experiments, forces us to abandon
this classical, local, way of describing the physical world. It has been
evident
for at least fifteen years that the quantum world is ruled by essentially
nonlocal mechanisms, and there is no way to reduce this nonlocality to
classical local objects. We propose that this fact of the physical world be adopted
as a fundamental physical principle.

Thus, we propose a fundamentally different mechanism of how events
affect each other. While the classical viewpoint is essentially local, we propose
that any two events (points) reflect events in each-other's vicinity,
 and this
is an immediate reflection without any notion of a signal traveling from one
point to another. This may sound absurd at first, but its full
implications are very naturally intuitive and consistent with observation.
This is because at any given point $x$, the reflections from other points
\textit{add up} to describe the events in the immediate neighborhood of $x$.
For example, the  value
of a field at a point is the sum of the values reflected from all the other
points of spacetime. Thus, even though each point reflects events everywhere
else, all points do not look the same. Also, as we look at successive 
space-like
sections, we perceive some effects to be moving from point to point and with 
definite speeds. Thus, although our hypothesis asserts a strong
\emph{action-at-a-distance,} its cumulative effects may
appear to be traveling from point to point with definite speeds;
consequently, the classical viewpoint
is not contradicted, but is supplemented at a more fundamental level. 
We call this
hypothesis of events being reflected in different neighborhoods the
nonlocality principle.

This principle can be refined using mathematical
language. For this we need to define two basic concepts. Following 
Einstein, we think of the universe as the set of events: each event 
corresponding to a mathematical point in spacetime\footnote{Actually it is
possible to formulate the principle---without any reference to a pre-existing 
spacetime---in the mathematical
language of categories; essentially by formalizing the way one arrives at the 
concept of a point from that of a neighborhood. It is not clear whether this 
will add to our understanding 
of physical phenomena, though.\label{categorical}}; but we call them 
\textbf{point events} instead, to underscore their exact conceptual content. 
Now we define
an \textbf{event} to be any set of point events. E.g., 
entire spacetime is an event,
and so is a single point event. Also, the trajectory 
(or
part thereof) of an electron is an event. Now we formulate the
\textbf{nonlocality principle} more precisely:
\par
\textit{Spacetime, $\mathrm{X,}$ the set of point events, is a
four-dimensional smooth manifold, such that
every pair of point events, $\mathrm{(x,y)},$ is nonlocally connected 
in the
following sense: given a pair of points $\mathrm{x}$ and 
$\mathrm{y},$ and their neighborhoods $\mathrm{U}$ and $\mathrm{V}$, respectively,
then events in each one of the
neighborhoods $\mathrm{U}$ and $\mathrm{V}$, are
reflected onto the other, $\mathrm{x}$ and
$\mathrm{y}$ being images of each other: these reflections are described by smooth onto maps between
neighborhoods, and are asymptotically exact in the following sense. As the
neighborhoods become smaller, the reflections converge to inverse, one-to-one,
reflections.}

\par
The consequences of the principle will follow
regardless of how we choose to formalize the asymtotic convergence (there are
several ways); so we can afford to be vague about the latter---at least for the
time being.

We can 
deduce entire 
physics from this
single hypothesis. In particular, we propose that nothing exists but this
scheme of reflections between pairs of neighborhoods. Thus, our theory does not even
assume the existence of matter, energy, fields, particles etc.; rather,
we deduce all these from the nonlocality principle as formulated above. 

The seminal consequence of the principle is that it implies a nonlocal
connection on spacetime. We use the word `connection' in the sense that it lets
us compare vectors at any pair of points in spacetime. The classical
connection, as conceived in differential geometry, is \textbf{local}
in the sense that it is essentially a way of relating vectors at any point
with those in its immediate (infinitesimal) vicinity.
\textit{A posteriori,} it allows us to compare vectors at distinct  points
through parallel transport along paths. We note that this way of
comparing vectors at distinct points depends on the path along which one
transports the vectors. This again points out the fact that the classical 
connection is essentially an infinitesimal object whose integral is the 
classical parallel transport.
As opposed to that, our nonlocal connection is
\textbf{nonlocal} in the sense that it provides a means of comparing vectors at
any two distinct points \emph{directly}, without any primary notion of 
infinitesimal transport of vectors or the accompanying
path-dependant parallel transport. We will also show that this nonlocal
connection has an infinitesimal (local) aspect which is nothing but a
classical (local) connection. Our theory accomplishes three
important objectives:

(1) The nonlocal connection \emph{explains} all
quantum aspects of the physical world.

(2) The local, infinitesimal, aspect of this connection gives a \emph{unified} description all the 
fields,
particles, quantum numbers, charges, mass, energy, momentum, etc.

(3) It is devoid of singularities.

We observe that the quantum
aspects are more apparent at smaller scales because the nonlocal
reflections grow more and more accurate as the neighborhoods grow smaller. Notions
of
`rate' at which this convergence to perfect accuracy occurs depend on notions of
`size' of neihborhoods . The latter depend on the
metric, the origins of which will be examined in the body of the work.

We will not go
into this any farther because we can formulate our theory without any
reference to it, or any other constants---dimensional or dimensionless.

If an intuitive picture of the universe is sought, we can say that
it is a giant kaleidoscope, each point being an infinitesimal mirror 
reflecting all other mirrors. Another
metaphor would be a cross-section of a bundle of optic 
fibers, which are fused 
together at 
one end into 
a single point. Yet another visualization of these nonlocal connections is
the image of a telephone exchange, where each point of spacetime
corresponds to a telephone; each phone being in direct instantaneous
communication with all the other phones. Then, cumulative information at each
point may appear to travel at finite speeds despite the underlying 
instantaneous communication among the phones.

\subsubsection{Preliminary consequences of the principle} 
\paragraph{Nonlocal connection} As a consequence of the principle,
there exists a one-to-one correspondence between vectors at $x$ and vectors at
$y$. It is easy to visualize this correspondence. Every vector can be thought
of as a 
tangent to a particle trajectory. Since events, such as particle 
trajectories, are 
reflected between pair of neighborhoods, we see that this induces a 
correspondence of 
vectors at points in these neighborhoods. Mathematically, this 
correspondence is an isomorphism from the
tangent vector-space $T_{x}$ at $x$ to the tangent vector-space $T_{y}$ at
$y$; roughly speaking, this isomorphism, say $\lambda_{xy},$ 
is
the `derivative'  or an infinitesimal limit of the correspondence referred to in the nonlocality 
principle.
Thus, we have, for every pair of points in spacetime, a way to compare
vectors at one of the points with those at the other. This is reminiscent of
the notion of
connection from differential geometry, which lets us compare vectors at
a point with
those at points in its infinitesimal neighborhood. This, the classical kind of
connection, is consequently a \emph{local} connection. As opposed to that,
what we have above is best described as a \textbf{nonlocal} connection, say
$\lambda$, whose value \emph{at} an ordered pair of points $(x,y)$ is the
isomorphism $\lambda_{xy}$. Note that $\lambda_{xy}$ and
$\lambda_{yx}$ are inverse isomorphisms. Also, $\lambda_{xy}$ naturally
extends to the whole tensor algebra at $x.$

\paragraph{Sum of reflections}
Consider a fixed point $x$ in spacetime. For any other point $y$,
events around $y$ will be reflected in events around $x$. For example, if an
elementary
physical field $F$ takes the value $F(y)$ at $y$, then it will
contribute $\lambda_{yx}(F(y))$ to the value of the field $F$ at $x$.
Here $\lambda_{yx}(F(y))$ is the value of the vector $F(y)$ under the
map $\lambda_{yx}$. Thus, the field value at $x$ is the \textit{sum} of
all these contributions as $y$ ranges over entire spacetime.
This sum is described mathematically by an integral:
\setlength{\fboxrule}{.2pt}\setlength{\fboxsep}{12pt}
\begin{equation}
\fbox{$\displaystyle F(x) = \int \lambda_{yx}(F(y))dy$} \label{sum}
\end{equation}
Note that the integrand is a function on $X$ with values
in the vector-space
$T_{x}$. This integral is defined using a volume-form on $X,$ integrating vector-valued 
functions 
component-wise.  Volume-form of a manifold is determined only up to a scalar
multiple. We will see later that a metric determines a volume-form \emph{and}
the physical measurements of the components of any field are also dependant on this
the metric. These dependances compensate for each other, and the above integral
equation is unambiguous, i.e. independant of the metric.
From this equation, we see that, elementary physical fields are extremely nonlocal objects in the 
sense that values at each point depend on values at all other points---not just nearby
points.

\subsection {\textbf{Quantum consequences}}
\subsubsection{Two-slit experiment}
\label{duality}
Because of equation (\ref{sum}) we can view an elementary particle as a field
which 
may appear
localized in a portion of spacetime and yet be spread-out over entire
spacetime; e.g., we can visualize an electron as a very intense region 
of a
field. Now, since this part of the field is the sum-total of the field
everywhere else (see equation \ref{sum}), it can also be viewed as spread-out
over entire spacetime. Reciprocally, this nonlocal
summation can give rise to intense regions in the field, which are dependant
on the values of the field everywhere else. Thus, discreteness and
contiguousness exist simultaneously, and yet in a non-contradictory way. Also,
more localized the particle-like phenomenon is, more it comes under the purview
of nonlocal connection, and more it manifests its wave-particle duality. This
is the basic picture to keep in mind when trying to understand quantum
phenomena. The two-slit experiment becomes immediately comprehensible from 
this
picture. As an aside, we mention that since clumpiness naturally arises from 
the
nonlocal character of spacetime, it may explain COBE-type data and
distribution of galaxies.

\subsubsection{Quantum randomness}
Consider the history of an observer in time as a one-parameter family of
space-like sections of spacetime. Then, given a field, the nonlocality
principle implies that the observer will not be able to predict
exactly how the field will change over his own history: The fields need not be
smooth, or even continuous, thus, the usual calculus is powerless to capture the (essentially nonlocal)
relationships among fields, and we can't predict anything using this classical concepts.
 Since we
recognize fields and particles as the same entity, we see that it is not
possible to predict any event exactly as conceived in classical physics.
Thus, the observer is left with the feeling that the events are purely
random, and he is led to believe that physical objects such as particles
don't have physical properties until they are observed. All the famous
paradoxes of quantum theory are based on this assumption and on the undue
significance that the process of measurement receives due to it.

\subsubsection{Aspect-type experiments}
The basic picture mentioned above also makes Aspect-type results transparent,
lending a solid physical explanation for the violation of Bell's inequality.

\subsubsection{Quantum tunneling}
This is just a manifestation of the apparent randomness and unconnectedness of
two events: vanishing of a particle at one point, and its reappearance
elsewhere. The point is that a particle doesn't have to go \textit{through}
a wall to appear on the other side. The electron may not even have a contiuous world history, i.e.
it may not have trajectory in the classical sense of the word.
The field configuration over the whole
spacetime, when viewed as space-like slices, appears to evolve in such a way
that it exhibits nonlocal effects, such as presence of its particle on
one side of the wall in one instance, and on the other side in the next
instance.

\subsubsection{Deterministic choice}
We have already noticed that the nonlocality principle is an expression of an
extreme form of determinism. Despite this, there is considerable room for 
an illusion of choice in this theory. Consider the case of an elementary field being 
monitored by an
observer. At any instance in time according to his frame of reference, the
field configuration in his past is already determined. Taking into account the
total nonlocal dependence of the field, one would think that the field 
configuration in the future, too is completely determined. This is not the 
case: Since the value of the field at any point is given by an 
\textit{integral} over $X,$ there can be infinitely many configurations, which can be related to the past of
our observer by equation (\ref{sum}). Consequently, the 
future
of the configuration has a fair degree of freedom without the need to
alter the past. However, we note that any two of these configurations can differ only on a set of measure zero.
Thus, all these configurations are almost same. The word almost is used here in the strict mathematical sense of
measure theory. Thus given data on a part of the universe which has nonzero meausre, we can, in principle, always compute
the rest of the configuration modulo sets of measure zero. The difference on set of measure is not a serious
problem because actual measurements at a point can not be made. What we measure is the average value over
a small region containing the point, and averages are same for two fields differring on a set of meausure zero.

\subsection{\textbf{The Local Field }}\label{local}
The nonlocal connection $\lambda$ gives rise to a classical, local
connection in the following manner.
Consider a path $\gamma$ connecting two distinct points 
$x$ and $y$. Now we describe an
integration procedure similar to usual Riemann integral from calculus. Let a
finite sequence, $s$, of points $x_{0},\ldots,x_{n}$ on $\gamma$ be such that
$x_0 = x$, $x_n = y$, and $x_i$ is between $x_{i-1}$ and $x_{i+1}$, for $i =
1,\ldots,n-1$. This sequence determines a sequence of isomorphisms
$\lambda_{x_{i}x_{i+1}},\quad i = 0,\ldots,n-1.$ Let $ \lambda^{s}_{xy}$ be 
the 
isomorphism obtained by composing successive maps in
this sequence. Now, letting  $n$ tend to $\infty$ yields a 
map,
which we denote by the symbol $\lambda_{xy}^{\gamma}$. This map
depends on the path joining the two points $x$ and $y$, and defines a
parallel transport on $X$. The corresponding (classical, local) affine
connection be denoted by $\omega$. Note that this is a connection in the
bundle $A(X)$ of affine frames of $X$.

In classical realm, $\omega$ is our fundamental field
potential, and its curvature $\Omega$ the fundamental field strength;
various physical objects will turn out to be aspects of this field.
Note that $\omega$ is the only structure we have assumed for our 
spacetime. (Actually it is just the classical aspect of  our nonlocal connection $\lambda.$)
In particular, we have not assumed a particular metric tensor on the
spacetime $X$. We will see that all matter, energy, spin, helicity, 
charges, fields and
particles are aspects of this field; and this field is not something
arbitrarily sewn together from these constituents. Instead, it is a geometry
arising out of the nonlocal character of spacetime---the constituents are mere
aspects of this geometry.

\section{THE FIELD}\label{field}
\subsection{\textbf{Phenomena and Observation: Classical and Quantum}}
When we measure fields, each measurement is an average
over small region of spacetime, and as we have seen in (\ref{principle}), the effects of nonlocality
tends to decrease as the regions become larger. Thus the behaviour of fields measured at larger scales, i.e. if each measurement is performed of a large enough region
then the measured fields appear local,
i.e. classical. This classical
behaviour is fully described by partial differential equations, which we will deduce in the
next section (\ref{cp}).

As the sensitivity of measurement is increased, each measurement is an average over smaller region and the effects
of nonlocality become more apparent
and the fields (as measured by us) don't satisfy the partial differential equations. Indeed, they need not be smooth, or
continuous, even.
Thus, the nonlocal character of the universe
forces us to abandon description in terms of partial differential
equations and
adopt some nonlocal concepts for description, so we would be able to compute
empirical predictions. Furthermore, in a nonlocal universe, it is only
natural that laws of physics are best formulated in terms of nonlocal concepts.
For these reasons, we introduce nonlocal calculus and nonlocal differential 
equations
in the section (\ref{qp}). We then formulate the fundamental nonlocal field 
equation. This equation contains lot more information on our local fields than does
the local equation. Note that fields analyzed here are local but the analysis
is nonlocal. Note that this analysis rests on the assumption that we know the nonlocal
connection $\omega.$

Our final and ultimate task is to find the ultimate unknown, the nonlocal connection $\omega$
which is a truely nonlocal field. To find this field, we need to develop a calculus of such
truly nonlocal fields. Then, we introduce the notion of nonlocal differential equations, and
point out that $\omega$ automatically satisfies an identity, a nonlocal differential equation.
This we called \textit{the field equation.} All these matters are presented in the
section (\ref{sp}).

\subsection{\textbf{Classical Phenomena: Classical Observer}}\label{cp}

\subsubsection{Lorentzian metric and time}\label{lorentz}
We re-iterate the fact that we have not assumed a metric for our spacetime.
It turns out that it is intimately related to
how we have classically chosen to measure various physical quantities; specifically, the
measure of time.

Consider the case of spacetime which is the vector space $R^4,$ without any structure assumed.
As Einstein argued, our measurement of distance depends on our notion of
simultaneity, and hence on how we measure time. Thus, given an observer, he
takes an event (which is a collection of point-events) and fixes that as his reference
to time mesurements. This event consists of his ``world history". All the measurements of 
distances are then based on this reference time, commonly known as his proper-time. The reason why he has
chosen to take \emph{time} as his reference lies in the way he experiences the world, i.e. the topology
of his consciousness. Besides this issue, there is no inherent reason for him to 
choose this collection of events, his proper-time, as his reference. It just happens to be convenient for
him to make this choice. As soon as he makes this choice, his space measurements are fixed. Now,
if there is another observer moving with respect to him, and making a similar choice of his notion of time,
his own space measurements are fixed. Becuase of this dependence, there are world-histories whose slope, say $c,$
remains unchanged as compared to that measured by the first observer. This slope is recognized as the
speed of light. Thus, both the observers are forced to deduce a Lorentzian metric being present.
Now consider a straightline which has a slope greater than $c$  (the speed of light)  as measured by one,
and hence, both of these observers. As we have seen, there is no inherent reason to exclude this
collection of events from being a reference for measuring other quantities. However, the constancy
of speed of light as observed by the first two observers mandates that this third ``observer" 
can not be observed by them. Also, from the viewpoint of the third observer, the first two can not be observed.
Furthermore, a fourth observer who \emph{can} be observed by the third one, will have his own notion of time
and the dependant notion of space, which will again isolate straightlines, whose slope, say $c',$ invariant between these two observers.
Again, they will deduce a Lorentzian metric.

Thus, if we choose to base our measurements on
fixing a notion of time and proceeding thenceforth, we see that the class of what we call ``frames of reference"
splits into mutually exclusive partitions; each partition containing those frames of reference, which are related to
each other by Lorentz transofrmations; and frames in separate partitions are mutually unboservable, i.e. 
can not be experienced as a time ordered sequence of events. Now, as soon as we choose one of these
frames of reference, we have limited ourselves to one such partition, and since only the frames of reference
in this partition are mutually observable, a consensus is reached among these frames of reference, that there
is an inherent metric to the spacetime. Thus, with each partition of frames of reference is associated one
Lorentzian metric, and it can easily be seen that different partitions assign different metrics to themselves in this
manner. Since the choice of frame of reference is an arbitrary one, we see that the metric, which is 
forced upon such choice is an abritary one.

Also, given any striaghtline, $l,$ we can find a partition, such that the slope of $l$ is an invariant among the
frames of reference belonging to this partition. Thus any `ray' is a `ray of light' for some class of frames of
reference, in that it can serve as the determining factor of Lorentz transformations which relate these frames of reference.
There is nothing inherently special about light as a physical phenomenon.

Now, this whole discussion, which pertained to $R^4$,
can be transfered to the tangent
spaces of individual point-events and the (infinitesimal) frames of reference at each such point-event.
Thus, as soon as we fix an infinitesimal frame of reference at any given point, we restrict ourselves to
a partition of (infinitesimal) frames of reference related to each other by Lorentz transformations; and to these frames of
reference, it would appear as if there is an inherent metric at the point-event. This holds for each point-event;
consequently, given an actual (not infinitesimal) frame of reference, which covers a region in spacetime, there is a
corresponding, automatic choice of infinitesimal frame of reference, and the corresponding metric,
at each point-event in this region. Thus, it would appear to this frame of reference that there is an
Einsteinian metric in the region of spacetime it covers. Consequently, we see that an Eisnteinian metric
is forced on us as soon as we choose a frame of reference, and the fact that \emph{it is Einsteinian} is forced on us becuase
of our choice of basing measurements on a time-based reference. Thus, we have an actual, physical,
reduction of the structure group from the general linear to Lorentz group.  As soon
as we choose to observe the universe in a time-bound fashion, the observable universe adopts this reduction,
this reduced amount of symmetries makes us believe that various aspects of the linear aspect of the local field $\Omega$
are invariant under this group, and forces us to recognize these split aspects as something inherent in our universe.
These splittings \emph{are}
various myriad froms (fields-particles) as observed
by the observer (and any other observer in his partion). 
We will study these splittings and the accompanying equations in the next section. For now, we note that these splittings 
and the equations are independant of this arbitrary choice of metric,
i.e. the chosen reduction of the general linear group to Lorentz group.

Now, we note that any closed curve in spacetime is time-like with respect to some metric. Thus, even in noncompact
universe, one can not escape the possibility of closed time-like curves. Consequently, it would appear that universe
is noncausal and that this noncausal character is dependant on the metric chosen. This forces us to look at the notion
of causality more closely. We start with a concrete example. Consider a classical particle with definite properties at a certain
point in spacetime, and having a closed world-line, which can be considered time-like according to some
metric. As it approches its initial position after
traversing it's closed path, if the values of it's characteristic properties aproach a value different than the one at the
initial position, then there will be a break of causality. Thus, the break in causality is equivalent to the mathematical
concept of discontinuity. Since all our fields (and hence particle trajectories), are smooth, and hence continuous, this
circumstance can not arise and the aforementioned properties of the particle, as it approaches the initail position,
will be forced to approach the same as those actually present at the intial position. The same can be argued with more
complex objects, such as human beings. Thus, our universe, although it permits
``closed timelike paths", does not lead to break-down of causality in classical picture.

Let us consider the integral (\ref{sum}) and the ambiguity in choice of the volume-form used in that integral.
We first notice that given a Lorentzian metric, there is a corresponding volume-form. Thus the volume form
depends on our choice of an oberver, specifically, on the notion of time of this observer. On the other hand,
our measurements of the components of field $F$ are also dependant on the notion of time of this observer.
Following Einstein's considererations on how space measurements get contracted upon change of observer,
it is easy to see that all measurements based on time get contracted upon such change. This contraction
compensates exactly for the multiplication by a scalar to the volume-form that such change entails. Thus, there
is no ambiguity in the definition of the integral in question.

Finally, we notice that our choice of observing the universe in a local, time-bound manner, forces us to see only
the local, i.e. point-wise, aspect  $\Omega$ of the nonlocal field $\Omega .$ However, since we can not escape the nonlocality, this field is
observed as compounded by the integral (\ref{sum}).

\subsubsection{The local field equation}
The bundle $A(X)$ of affine frames on $X$ is a
principle fiber bundle with structure group $GA(4, \mathbb{R})$, the general
affine group, i.e. the full
group of affine automorphisms of the four-dimensional real affine
space $\mathbb{R}\mathnormal{^4}$. The affine connection $\omega$ is represented by a 
$1$-form on $A(X)$
with values in the Lie algebra $\mathfrak{ga}(4, \mathbb{R})$ of the Lie group
$GA(4,\mathbb{R})$. We denote this $1$-form by the same letter $\omega$. If
$D$ denotes the covariant exterior derivative with respect to
$\omega$, then the Bianchi identity,
$D^2\omega = 0,$
holds.\label{ie}
We define \textbf{curvature} $\Omega$ of a connection $\omega$ to be the
covariant exterior 
derivative
of $\omega$ with respect to itself: $\Omega = D\omega.$
Thus, Bianchi identity can be rewritten:
\begin{equation}
\fbox{$\displaystyle D\Omega = 0$} \label{bianchi}
\end{equation}
This we call the \textbf{local field equation.} Rather than being an additional
hypothesis,
the local field equation is a consequence of the fact that $\Omega$ is a
curvature. Note that this equation is a conservation principle. Also, 
since $D$ depends on $\omega$, the field equation is
\emph{non-linear}.

\subsubsection{Forces and Charges}

\paragraph{Indentifying  forces and charges}
Now, the bundle $A(X)$ naturally contains the bundle $L(X)$ of linear frames on
$X.$ The Lie algebra of the affine group splits naturally; 
$\mathfrak{ga}(4,\mathbb{R}) = \mathfrak{gl}(4,\mathbb{R}) \oplus 
\mathbb{R}\mathnormal{^4},$ where the first term on the right is the Lie algebra of the
general linear group. Corresponding to this split,
the restriction of $\omega$ to $L(X)$ splits into two forms,
$\omega = \lambda + \theta$
where $\lambda$ is $\mathfrak{gl}(4,\mathbb{R})$-valued, and $\theta$ is the
$\mathbb{R}\mathnormal{^4}$-valued \textit{canonical} form of the bundle $L(X).$
The curvature of $\lambda$ is
$\Lambda = D\lambda,$
and the \textbf{torsion} $\Theta$ of the connection 
$\lambda$ is defined by
$\Theta = D\theta,$
$D$ being with respect to $\lambda.$ Then, on $L(X),$ the following holds:
$\Omega = \Lambda + \Theta.$
Due to this \ref{bianchi} implies the following
Bianchi identities for $\lambda$:
\begin{equation}
\fbox{$\displaystyle D\Lambda = 0,\hspace{6pt}D\Theta = \Lambda \wedge 
\theta$}\label{bianchi4}
\end{equation}
Here the covariant exterior derivative $D$ is with
respect to $\lambda.$

The 2-forms $\Lambda$ and $\Theta$, the linear and rotational aspects of 
$\Omega,$
respectively, contain all the physical fields along with
various quantum numbers such as charges, spin, helicity etc. We will
see next how to identify this information in these forms. Meanwhile,
we notice that these two forms are mathematically equivalent to tensor
fields $L$ and $T$, respectively, on the base space $X$. We call $\omega$, and
$\Omega$, $\Theta$ \textbf{universal local connection}, \textbf{universal local curvature,}, and
\textbf{universal local torsion form,} respectively. The last two are equivalent to two
tensor fields $U$ and $T$ on $X,$ and are called \textbf{universal local field} and
\textbf{universal local torsion,} respectively. These satisfy $U = L + T.$

We identify the trace $\overline{T}$ of $T$ with the
spin-helicity vector: the three space components of $\overline{T}$ representing
the spins in three space directions, while the fourth, the time
component, representing the helicity. Together, we call these the components of
the \textbf{spin-helicity} field.  The traceless symmetric part
$\underline{T}$, on the other hand, represents new properties of particles
that we call \textbf{spun} and \textbf{heluxity}. \label{spin-n-stuff}

Now, given an observer, as described in (\ref{lorentz}), our 
frame bundle $L(X)$ physically reduces to a Lorentz subbundle $G(X)$, with
the Lorentz group $G.$  as the structure group.\label{reduction} (actually this happens only on the region covered by
the observer, but we still use the letter $X$ for this region.)
The general linear algebra splits as a direct sum
$\mathfrak{gl}(4, \mathbb{R}) = \mathfrak{g}\oplus \mathfrak{d},$
where $\mathfrak{g}$ is the Lie algebra of the Lorentz group, and
$\mathfrak{d}$ is a subspace (not a Lie subalgebra) invariant under the
restriction of the adjoint representation of $GL(4,\mathbb{R})$ to the Lorentz
group. Indeed, $Ad(G)(\mathfrak{d}) = \mathfrak{d}$.
Because of this, the restriction of $\lambda$ to $G(X)$ splits 
into
two forms
$\lambda = \gamma + \delta ,$
with $\gamma$ being a connection on the principle bundle $G(X)$ (not the 
Levi-Civita connection), and $\delta$ is a $\mathfrak{d}$-valued
$1$-form invariant under the adjoint action of $G$ on $\mathfrak{d}$. 
Corresponding to this split, the curvature form $\Lambda$ splits into two
components,
$\Lambda  =  \Gamma + \Delta,$
where
$\Delta = D\delta,$
the exterior covariant derivative of $\delta$ with respect to the
connection $\gamma$. Note that $\Gamma$ is $\mathfrak{g}$-valued, whereas 
$\Delta$ is $\mathfrak{d}$-valued. These forms are equivalent to $2$-forms on
the base space $X$, with values in a $\mathfrak{g}$-bundle and a 
$\mathfrak{d}$-bundle
respectively, associated with the principle bundle $G(X)$. These in turn are
equivalent to tensor fields $G$ and $D$
so that the curvature tensor $L$ of $\lambda$ splits as
$L = G + D.$
Here, $G$ is the Riemann curvature tensor field of the metric
connection $\gamma$, (this is not the Riemann curvature tensor of the
Levi-Civita connection), whose value at a point $x$ is in the tensor
space $T_{x}\otimes T_{x}^*\otimes T_{x}^*\otimes T_{x}^*$. We will see that
this tensor contains the energy-momentum tensor as well as the
gravitational field. Note that the above split spells the exact
relationship between gravitational aspects, $G,$ and non-gravitational 
aspects, contained in $D.$

The adjoint representation of $G$ described above splits into
irreducible component representations,
$\mathfrak{d} = \mathfrak{f} \oplus\mathfrak{p},$
where $\mathfrak{p}$ is the subspace consisting of all traceless, symmetric
$4\times4$ matrices, and $\mathfrak{f}$ is the subspace consisting of all the
scalar multiples of the identity matrix, $diag(1, 1, 1, 1)$. The
corresponding $Ad(G)$-invariant decomposition of $\Delta$ is
$\Delta = \Phi + \Pi.$ Correspondingly, $D$ splits into two tensors on the base
space: $D = F + P.$
Later, we will see how $F$ contains the electromagnetic phenomena (the
force field and the charge), and $P$ turns out to be the tensor product of the
strong and the weak phenomena. Indeed,  $\mathfrak{p}$ is a tensor product of
two irreducible $Ad(G)$-invariant components,
$\mathfrak{p} = \mathfrak{s} \otimes\mathfrak{w},$
with corresponding decompositions of the $2$-form $\Pi$ and the field $P$,
$\Pi = \Sigma \otimes \Psi,$ and $P = S \otimes W.$

Thus, we have derived three more fields, $F,$ $S,$ and $W,$ besides the Riemann
curvature tensor $G$.

To summarize, we collect:
\begin{gather}
\Omega = \Gamma + \Phi + (\Sigma\otimes \Psi)+ \Theta\\
U = G + F + (S \otimes W) + T \label{Usplit}\\
\fbox{$ \displaystyle D\Gamma + D\Phi + D(\Sigma \otimes \Psi) + D\Theta = 0$}
\label{detailfe}
\end{gather}
When all but one of these fields are assumed absent, we have, respectively,
\begin{equation}
D\Gamma = 0,\hspace{3mm}
D\Phi = 0,\hspace{3mm}
D(\Sigma\otimes\Psi) = 0,\hspace{3mm}
\end{equation}
and will be called the local \textbf{gravidynamic,
electrodynamic,} \textbf{chromoflavodynamic} 
equations respectively. Here, $D$ is with respect to $\gamma, $ the linear connection.
Of course, several other equations can be deduced
under other
conditions.

Consider the Riemannian curvature tensor $R.$ It has two aspects: $\underline{R},$ the Weyl conformal tensor,
and $\overline{R},$ the Ricci curvature tensor. These are essentially the ``traceless" and the ``trace"
parts of $R,$ and completely determine the latter.  More generally, each of our fields, $G, F, S, W,$
and $T$ derived from $\Omega$ can be analyzed into traceless and trace parts: $\underline{G}, \overline{G}; 
\underline{F}, \overline{F}; \underline{S}, \overline{S}; \underline{W}, \overline{W}; \underline{T}, \overline{T}.$
We indentify the trace part with a source and the traceless part with the corresponding ``force" field.
Thus, for example, $\underline{G}$ and $\overline{G}$ are identified as the gravitational field and the
energy-momentum tensor. 

\paragraph{Conservation of forces and charges}
Note that the individual equations, gravidynamic, electrodynamic, and chromoflavodynamic,
are conservation laws. Since the operation (contraction) used in forming the trace part and the traceless part
of $G, F, S, W,$ and $T$ commutes with the covariant derivative, the above mentioned equations
give rise to similar equations for these individual sources and forces. Thus, for example, when any one of this equation is
satisfied,
i.e. when any one of $G, F, S, W,$ and $T$ is conserved, the corresponding source and supply are conserved individually.
Thus, for example, if  $D\Gamma = 0,$ holds, as it does in the absence of other fields, it follows that
\begin{equation}
D\underline{G} = 0\hspace{3mm}, D\overline{G} = 0\hspace{3mm}.
\end{equation}
Similar equations hold for the remaining four fields.
The trace and traceless parts of the remaining fields $S, W,$ and $T$ are identified  respectively with  strong (color)
current density and strong force fields; weak (flavor) current density and weak force fields; spin-helicity field and the correspondoing
``force" field, spun-heluxicity.
Thus, it is seen that sources and forces are just aspects of a unified whole.

\paragraph{Einstein and Maxwell equations}
When all other fields are assumed absent,
the identification of $\overline{G}$ with the energy-momentum tensor
reduces to the Einstein equation, since energy-momentum tensor is necessarily conserved.
Similarly, the identification of $\overline{F}$ with the current density, and of $\underline{F}$ with the electromagnetic
field implies the Maxwell equations.
\label{source-supply}

\subsubsection{Structure of particles and interactions}
\textit{\newline Basic constructions:} 
We propose that the observed particles/fields are nothing but the
manifestations of the field $\Omega$ with varying intensity of the
constituent fields. For instance, electron is a field/particle whose only 
nonzero
components are (i) negative charge field (ii) energy-momentum field (iii)
spin-helicity field.  Similarly, we view leptons, quarks, and
so-called gauge bosons as particles/fields with various combinations of
nonzero field components.
\newline
\textit{Interactions:} 
All the elementary fields interact \emph{because} they have to satisfy equation 
(\ref{detailfe}).  All the
interactions are built into this equation.

\subsection{\textbf{Quantum Phenomena: Classical Observer}} \label{qp}
\subsubsection {Nonlocal calculus of fields}
\paragraph{Nonlocal integration}
With the terminology of (\ref{sum}), the \textbf{nonlocal
integral of any field $F$ at $x$,} $(\int F)(x),$ is defined by
\begin{equation}
(\int{F})(x) = \int{\lambda_{yx}(F(y))dy}
\end{equation}
This defines
a new field $\int F,$ called the \textbf{nonlocal integral} of $F$. Note that $F$ 
encodes the nonlocal dependence of $\int F.$ Thus, an important question 
arises: for
what field(s) $f$, $F = \int{f}$? In other words, we would like to know how 
a field is
nonlocally put together. For this we define nonlocal derivative.

\paragraph{Nonlocal derivative} 
The \textbf{nonlocal derivative} of a field $F$ is defined to be a field $DF,$ given
by 
\begin{equation}
DF(y) = \lambda_{xy}(\partial_{y}{F}_{x}(F(y)).
\end{equation}
Where $F_{x}$ is the function given by $F_{x}(y) = \lambda_{yx}(F(y)),$ 
and $\partial_{y}{F}_{x}$ stands for its derivative at $y.$ The 
significance
of the nonlocal derivative is that the following theorem holds.

\paragraph{Fundamental theorem of nonlocal calculus}
\newtheorem{theorem}{Theorem}
\begin{theorem}
Let $F$ be a field, then
\begin{equation}
\fbox{$\displaystyle \int{DF} = F = D\int{F}$}
\end{equation}
\end{theorem}
This theorem establishes the crucial conceptual link between (nonlocal)
derivation and (nonlocal) integration. Also, several 
computations that would be impossible to carry out are made possible by this
theorem.

Following the case of (local) derivatives, we can repeat the process of
taking nonlocal derivative and get successive derived fields $F^{(1)},...,
F^{(n)},...$ Note that $\int(F^{(n)})$ is $F^{(n-1)}$. Here
$F^{(n-1)}$ prescribes the nonlocal dependence of field $F^{(n)}$. Also, we
write $F^{(-1)}$ for $\int F$ and by repeated application of the operator
$\int$, we get a 
sequence of
fields which we denote by $F^{(-2)}, F^{(-3)},\ldots$ etc. Thus, we can talk
about fields $F^{(n)}$ for any integer $n$ (with the convention that $F^{(0)}$
denotes $F$). Now we are ready to formulate nonlocal equations. 

\subsubsection{Nonlocal field equation}
\paragraph{Differential equations}
A \textbf{nonlocal differential equation} in terms of an unknown field $F$ is
one which involves functions of $F^{(n)}$ where $n$ may take finitely many
integer values. Symbolically,
\begin{equation}
\mathcal{G}\mathnormal{(F^{n_1},...,F^{n_k}) =\mathbf{0,}}
\end{equation}
where $k$ is a positive integer, and $\mathcal{G}$ is a function 
of $k$ arguments.
With this vocabulary, the nonlocality principle implies that an elementary 
field
satisfies the fundamental nonlocal equation
\begin{equation}
dF = 0. \label{nonlocal}
\end{equation}
where $dF = F - DF.$ Note that this equation is just a re-wording of
equation (\ref{sum}).

All fields measured at very small scale satisfy this nonlocal equation. This is not true for
other 
physical fields, such as heat and sound in gases. This is because the
phenomena 
concerned are generally scaled at macro scales at which the effects of 
nonlocal 
connections are not significant. 
But when we investigate the behavior of fields at scales where nonlocality effects are significant,
they satisfy equation (\ref{nonlocal})---and no partial differential equations are satisfied. Indeed, measured
at this scale, fields can not even be assumed smooth. Thus, our only recourse in this case, is to solve the
nonlocal equations.

\paragraph{Nonlocal field equation}

To begin with, we note that the universal local field $U,$
satisfies the nonlocal differential equation
\begin{equation}
\fbox{$\displaystyle dU = 0$} \label{nfe}
\end{equation}
This equation
will be called the \textbf{nonlocal field equation.} Indeed, (\ref{nfe}) 
contains enough information to let us
extrapolate partial data on the field configuration to entire spacetime. Also,
it is a consequence of nonlocality principle rather than being an additional
hypothesis. Now we derive field equations for various field constituents of 
the universal
field $U.$ From (\ref{nfe}) and (\ref{Usplit}) we have
\begin{equation}
\fbox{$\displaystyle dG + dF + d(S\otimes W) + dT = 0.$} \label{eq3}
\end{equation}
Assuming absence of all but one of $G, F, S, W, T,$ at a time, we get
\begin{equation}
dG = 0,\hspace{3mm}dF = 0,\hspace{3mm}d(S\otimes W) = 0\hspace{3mm}dT = 0,
\end{equation}
respectively. We call these equations
the nonlocal \textbf{gravidynamic}  equation,
the nonlocal
\textbf{electrodynamic} equation, and the nonlocal \textbf{chromoflavodynamic}
equation,
respectively.

\subsection{\textbf{Quantum Phenomena: Quantum Observer}} \label{sp}
The most fundamental structure on spacetime is the nonlocal connection 
$\lambda.$
This is an example of what we will call a (nonlocal) form. Unlike the usual
fields, this field is defined at an \emph{ordered pair of points,} and is an 
isomorphism 
from the tangent space of the first point to that of the other. In order to
analyze $\lambda,$ we 
develop
an analysis of such forms.

\subsubsection{Calculus of nonlocal fields}

\paragraph{Chains and Forms}

We start with several definitions related to the concept of a
\textbf{$p$-chain.}
For a positive integer $p,$ the $p$-cube $I^p$ is the set $[0,1]^p$ in
$\mathbb{R}^{p}.$ Also, $I^{0}$ is just a singleton set, fixed once and for
all. The point
$(0,\ldots,0)$ in $\mathbb{R}^{p},$ is denoted by $e_0,$ and the
$i^{\text{th}}$ unit vector for
$i = 1,\ldots,p$ is denoted by $e_i.$ For $p = 0,1,2,\ldots$ a 
$p$-cube in a manifold $X$ is  map $C:I^{p}\rightarrow X.$
The point $C(e_i), i= 0,\ldots,p$ is called the $i^{\text{th}}$ vertex
of $C,$ and is denoted by $x_i.$ We call $x_0$ the initial vertex and $x_p$
the final vertex of $C.$ For the standard $p$-cube $I^p$ we define
faces $I_{(i,0)}, I_{(i,1)},$ for $i=1,\ldots,p,$ to be the $(p-1)$-cubes in
$\mathbb{R}^p$ given by $I_{(i,\alpha)}(x) = (x_0,\ldots,x_{i-1},\alpha,x_{i},
\ldots,x_{p-1}),$ where $x$ is any point in $I^{(p-1)},$ and $\alpha = 0,1.$
The faces of a cube $C$ in $X$ are $(p-1)$-cubes in $X$ given by
$C_{(i,\alpha)} = C\circ I_{(i,\alpha)}.$ A \textbf{$p$-chain} in $X$ is an
element of the vector space generated by the $p$-cubes in $X.$ The boundary 
$\partial$ of a $p$-cube $C$ is
the $(p-1)$-chain given by
\begin{equation}
\partial C = \sum_{i=1,\ldots,p ;\hspace{3pt}\alpha = 0,1}(-1)^{(i+\alpha)}
C_{(i,\alpha)}.
\end{equation}
The following proposition holds:
\begin{Proposition}
For any chain $C,$ we have $\partial^{2}C = 0,$ i.e.
\begin{equation}
\partial^2 = 0.
\end{equation}
\end{Proposition}

We inductively define an \textbf{edge} of the standard cube $I^p$ as follows.
An edge of $I^0 = {0}$ is just the ordered pair of points $(0,0).$ 
An edge of a $I^1$-cube is simply the ordered pair consisting of its initial
and final vertices in that order. An edge of $I^p,$ for $p>1$ is defined to be
any an edge of any of its faces. 
A \textbf{$p$-form} is an assignment to every $p$-cube $C$
of a set of affine homomorphisms of tangent spaces (at vertices) along its
edges. Now we notice that a nonlocal connection 
$\omega$ is just a $1$-form on $X,$ where the homomorphisms are actually 
isomorphisms. A $0$-form is an assignment of an affine operator on the tangent
space at each point.

\paragraph{Derivative}
Given a $p$-form $\omega,$  we define its \textbf{derivative} to 
be a 
$(p+1)$-form $d\omega$ given on a $(p+1)$-cube $C$ by applying $\omega$ on its
faces $C_{(i,\alpha)}$ with a multiplier $(-1)^{(i+\alpha)}.$ We formally write
this as follows:
\begin{equation}
(d\omega)_{C} = \sum_{i,\alpha}(-1)^{i+\alpha}\omega_{C_{(i,\alpha)}}
\label{derivative}
\end{equation}
We record the following:
\begin{Proposition}
For any form $\omega, d^2\omega = 0,$ i.e.
\begin{equation}
d^2 = 0.
\end{equation}
\end{Proposition}
\paragraph{Integral}
The \textbf{integral} $\int_{C}\omega$
of a $p$-form $\omega$ over a $p$-cube $C$ is a homomorphism from the
tangent space of its initial vertex to that of its final vertex, which is 
simply the sum of all the various compositions
of homomorphisms along the edges of $C$; each of these compositions begin
and end at the $T_x.$ This definition of integral can be extended to the
boundary of a cube
$C$ by first evaluating the integral over each face $C_{(i,\alpha)}$, 
multiplying it with the multiplier $(-1)^{(i+\alpha)}$ and then further
composing those from initial to final vertices of $C.$

Now we record the relationship between the integral and derivative in this
nonlocal analog of Stoke's theorem.
\begin{theorem}[Fundamental theorem of nonlocal calculus]
Given a $p$-form $\omega$ and a $(p+1)$-chain $C,$ we have the following:
\begin{equation}
\fbox{$\displaystyle \int_{\partial C}{\omega} = \int_{C}{d\omega}$}
\end{equation}
\end{theorem}

\subsubsection{Nonlocal differential geometry}
We consider a fiber bundle on the space $X^2.$ For every ordered pair 
$(x,y)$
in $X^2$ the fiber is the set $L(T_x,T_y)$ of homomorphisms from 
$T_x$ to
$T_y.$ This bundle will be 
denoted by $L$. 
The sub-bundle, $L(X^{2}),$ of $L$ consisting of isomorphisms
 has fiber $L(T_x,T_y)$ at $(x,y).$
Let a pair $(u,v)$ of frames, one for each $T_x$ and $T_y,$ be given; i.e., let
isomorphisms $u:\mathbb{R}\mathnormal{^4}\rightarrow T_x$ and
$u:\mathbb{R}\mathnormal{^4}\rightarrow T_y$ be given. Then we have
a unique
isomorphism $vu^{-1}:T_x\rightarrow T_y$ in $L(T_x,T_y).$ Now, the
general linear group acts on the set of pairs of such frames by 
$g(u,v) = (ug,v) = (u,vg^{-1}),$ where $g$ is an automorphism of 
$\mathbb{R}\mathnormal{^4}.$ It can be
easily verified that this makes $L(X^{2})$ a principle fiber bundle with 
structure group $GL(4,\mathbb{R})$. Now we see that a nonlocal connection
is 
nothing 
but a section
of the principle bundle $L(X^{2}).$

We define the \textbf{derivative} of  a
$p$-form $\psi,$ with respect to a connection $\omega,$ to be a
$(p+1)$-form 
$D\psi,$ given on a $(p+1)$-cube $C$ by the same formula (\ref{derivative}) but
replacing $\psi$ by $\omega$ on edges other than those 
on $C_{(1,0)}$ and on $C_{(1,1)}.$

We define the \textbf{curvature $\Omega$} of a nonlocal connection $\omega$ 
to 
be its
derivative with respect to itself, i.e. $\Omega = D\omega.$
A connection will be called \textbf{flat} if the curvature $\Omega$ vanishes identically.

We record the following theorem:
\begin{theorem}
The connection $\omega$ is flat if and only if the correponding infinitesimal connection $\omega$ is flat.
\end{theorem}

The following identity is the nonlocal analog of Bianchi identity from (local)
differential geometry.
\begin{Proposition}
The identity $D\Omega = 0,$ holds for any connection $\omega.$
\end{Proposition}

\subsubsection{The field equation}

Assuming that a Lorentzian metric is given
on $X,$ we have a reduction $G(X^2)$ of the principle bundle $L(X^2)$ with
structure group $G$, the Lorentz group. Correspondingly, $L$ splits into two
bundles, $L = G \oplus 
D.$ Where $G$ consists of homomorphisms preserving the metric, and 
$D$ is invariant under the action of the group $G.$.
Thus, $\Lambda$ splits into a
$G$-valued part $\Gamma,$ and its $D$-valued complement 
$\Delta$: 
$\Lambda = \Gamma + \Delta.$
Finally, $D = F \oplus (S \otimes W),$ where each of the three 
bundles on
the right are (group)$G$-invariant.
Hence,
$\Delta$ splits into three parts: 
$\Delta = \Phi + (\Sigma\otimes \Psi).$
Thus we have the decomposition,
\begin{equation}
\Lambda = \Gamma + \Phi + (\Sigma\otimes \Psi). \label{everything}
\end{equation}
This equation describes how $\Lambda$ naturally splits into its 
constituent nonlocal
fields: $\Gamma$ is the gravitational aspect, $\Phi$ the
electromagnetic aspect, $\Sigma$ and $\Psi$ are the strong and
weak aspects, respectively. All the forces and sources are contained in the same unified whole.
\footnote{Note that there is no mention of a nonlocal `torsion' $\Theta.$ This is because torsion is a purely
local concept and there is no nonlocal analog for it. Thus, at nonlocal level, we only analyze the
linear connection $\lambda,$ and the nonlocal curvature $\Lambda.$ Knowledge of $\lambda$
determines the local affine connection $\omega,$ and the local curvature $\Omega$ includes the
torsion of the local connection $\lambda.$}
Just as in the case of
local universal field, we take the trace part of these nonlocal fields and 
identify those with the sources of the traceless force fields. For example,
$\overline{\Gamma}$ and $\underline{\Gamma}$ are the
energy-momentum field and the gravitational field, respectively (see
Sec. \ref{source-supply}).
The identity above, when applied to our universal nonlocal curvature,
gives us \textbf{the field equation:}
\begin{equation}
\fbox{$\displaystyle D\Lambda = 0.$}
\end{equation}
This equation, along with (Eq. \ref{everything}), yields a detailed
field equation:
\begin{equation}
\fbox{$\displaystyle D\Gamma + D\Phi + D(\Sigma \otimes \Psi) = 0$}
\end{equation}

\section{THE CONCLUSION}\label{conclusion}

\subsection{Rotation of polarization of Light}\label{rotation}
The new field
\textbf{spun-heluxicity}, the traceless part accompanying
the field spin-helicity (\ref{spin-n-stuff}), would rotate the
polarization of light{\footnote{Just before the submission of 
this paper, it was pointed out to the author that such phenomena have 
already been observed.} (and other such `directed' properties) just like
curvature bends light.

\subsection{Bending of light near magnetars}
Since electromagnetism is a non-metric aspect of the universal
connection, given a stellar object, such as magnetar, with an immense magnetic
field that is
comparable in curvature to its gravitational field, bending
of light should be significantly different than that
predicted by general relativity.

\subsection{Parallelizability and shape of spacetime}
We first note that existence of the nonlocal connection implies that spacetime has a trivial tangent bundle, and is orientable.
We also note that nonlocality principle immediately implies that spacetime can not be compact (because image of it under
the maps assumed in the principle is always an open neighborhood). Triviality of the tangent bundle strongly indicates the
possiblity of the sapacetime being an open set of $R^4.$
The simplest open set, $R^4$ itself is our choice candidate. Note that this has nothing to do with the notion of size; the latter is dependant on metric. We are
are talking of the \emph{topology} of spacetime; and from this viewpoint, $R^4$ and the interior of the $4-$sphere are the same same
thing.

\subsection{No-singularities and big-bang}
\label{nosingularities}
The general conservation principle (\ref{bianchi4}) implies that if 
the gravitation
part of curvature increases, then the non-gravitational part will compensate 
for it. Consequently,
black-hole type situation can't lead to singularities.
This mechanism, which prevents singularities, can be
interpreted as a sort of anti-gravity; it is not an extra force of nature,
but is built into our theory by virtue of the grand conservation law 
(\ref{bianchi}).

The lack of absolute metric implies that there are no absolute notions of 
expansion and contraction of space. Thus, expansion is not an absolute
feature of the universe.  Given any point in spacetime, we can find a 
metric such that the sections spacelike with respect to this metric
will appear to converge to this point. Thus no point of spacetime is any more, or less,
exotic than any other point.
Since the universe is noncompact, in may be incomplete; the `point' to which spacelike sections seem to be converging may not even exist.
On the other hand, if the universe is complete, i.e. without holes, such as $R^4,$ then this apparent
convergence does point to an actual point of spacetime---again there is nothing special about this point.

\subsection{Causality}
Notion of causality, in Newtonian world depended on the assumption of the notion of absolute time.
After Einstein, it depended on the notion of time-like interval between point-events. This time-like
character is an absolute in Einsteinian world. There is no absolute time, but at each point, the directions
in which `time' can flow, depending on the frame of reference, are restricted (`future cone' and `past cone').
And these depended on the assumption that there is a (Einsteinian-Lorentzian) metric present.
And the notion of causality was formulated in this more relaxed yet precribed directions of flow of `time'.
Furthermore, the notion of causality was supported by the fact that there were two \emph{separate} classes
of directions in which time could flow. So much so that a direction's character of being `time-like' or `not time-like'
was called `the causal character'. Thus, the notion of causality in Einsteinian world is supported by the existence
of an absolute metric of Lorentzian
character.

We have seen that even the metric is arbitrary, and given any direction at a point event, we can find a metric
according to which `time' can flow in that direction, i.e. a metric such that the given direction is in the `future-cone'
(or `past-cone') of that point. Consequently, there are no prefrered directions of flow of time, i.e.
\emph{no notion reminiscent of time};
no notion of absolute past and absolute present. There is no absolute notion of causality.

That classical fields do seem to follow causal law with respect to each imaginable metric rests on the fact that
classical fields are smooth and continuous, and so are the trajectories of classical particles. The metric gives us a chance
to talk about causality, and the classical smoothness gives us classical causality---with respect to one and hence all Lorentzian metric.

Since the fields satisfying the equation (\ref{sum})), i.e. fields as measured at very small scales,
may not be smooth or even continuous, we would deduce a break of causality. However, this is an error,
since there is no absolute notion of causality.

\subsection{Micro predictions}
Corresponding to the field spun-heluxity, there is a new particle property,
which should be inferable from observation of rotation mentioned in section
(\ref{rotation}) 
in particle interactions, as well.

Our viewpoint also validates particles of other
fields such as sound and heat when these are determined at, and measured at, micro-scales, e.g.
in solid-state. More generally, we
predict \textbf{anyons} corresponding to \emph{any} conceivable physical field 
determined at extremely small scale.

\subsection*{Acknowledgements}

Professing: Deep gratitude to my mentor, Prof. Phillip. E. Parker, for
providing a first 
critical assessment of the paper and pointing out several fine points. Also,
many thanks to my friend, Irfan Siddiqui, for several helpful discussions.

\end{document}